\documentclass[conference]{IEEEtran}
\IEEEoverridecommandlockouts

\usepackage{url}
\usepackage{graphicx}
\usepackage[usenames,dvipsnames,svgnames,table]{xcolor}
\usepackage{hyperref}
\usepackage{bchart}
\usepackage{multirow}

\definecolor{linkblue}{RGB}{0,0,155}

\hypersetup{
   bookmarksnumbered,
   colorlinks={true},
   pdfstartview={FitH},
   citecolor={linkblue},
   linkcolor={linkblue},
   urlcolor={linkblue},
}

\newcommand{\todo}[1]{}

\definecolor{acolor}{HTML}{99ccff}
\definecolor{pcolor}{HTML}{f3a08c}
\definecolor{icolor}{HTML}{ffff66}
\definecolor{hcolor}{HTML}{e8e8e8}

\begin{document}

\title{Nanopublications: A Growing Resource of Provenance-Centric Scientific Linked Data}

\author{
\IEEEauthorblockN{Tobias Kuhn}
\IEEEauthorblockA{Department of Computer Science\\Vrije Universiteit Amsterdam\\The Netherlands\\\texttt{t.kuhn@vu.nl}} \and
\IEEEauthorblockN{Albert Mero\~no-Pe\~nuela}
\IEEEauthorblockA{Department of Computer Science\\Vrije Universiteit Amsterdam\\The Netherlands} \and
\IEEEauthorblockN{Alexander Malic}
\IEEEauthorblockA{Institute of Data Science\\Maastricht University\\The Netherlands} \and
\IEEEauthorblockN{Jorrit H. Poelen}
\IEEEauthorblockA{400 Perkins St\\Oakland, California\\USA} \and
\IEEEauthorblockN{Allen H. Hurlbert}
\IEEEauthorblockA{Department of Biology\\University of North Carolina\\Chapel Hill, USA} \and
\IEEEauthorblockN{Emilio Centeno Ortiz}
\IEEEauthorblockA{Research Programme on\\Biomedical Informatics, Hospital del\\Mar Medical Research Institute\\Universitat Pompeu Fabra\\Barcelona, Spain} \and
\IEEEauthorblockN{Laura I. Furlong}
\IEEEauthorblockA{Research Programme on\\Biomedical Informatics, Hospital del\\Mar Medical Research Institute\\Universitat Pompeu Fabra\\Barcelona, Spain} \and
\IEEEauthorblockN{N\'uria Queralt-Rosinach}
\IEEEauthorblockA{Department of Integrative Structural\\and Computational Biology\\The Scripps Research Institute\\La Jolla, USA} \and
\IEEEauthorblockN{Christine Chichester}
\IEEEauthorblockA{Datafair.xyz\\Geneva, Switzerland} \and
\IEEEauthorblockN{Juan M. Banda}
\IEEEauthorblockA{Department of Computer Science\\Georgia State University\\USA} \and
\IEEEauthorblockN{Egon~Willighagen}
\IEEEauthorblockA{Department of Bioinformatics\\BiGCaT, NUTRIM\\Maastricht University\\The Netherlands} \and
\IEEEauthorblockN{Friederike Ehrhart}
\IEEEauthorblockA{Department of Bioinformatics\\BiGCaT, NUTRIM\\Maastricht University\\The Netherlands} \and
\IEEEauthorblockN{Chris Evelo}
\IEEEauthorblockA{Department of Bioinformatics\\BiGCaT, NUTRIM\\Maastricht University\\The Netherlands} \and
\IEEEauthorblockN{Tareq B. Malas}
\IEEEauthorblockA{Department of Human Genetics\\Leiden University Medical Center\\The Netherlands}  \and
\IEEEauthorblockN{Michel Dumontier}
\IEEEauthorblockA{Institute of Data Science\\Maastricht University\\The Netherlands}
}



\maketitle

\begin{abstract}
Nanopublications are a Linked Data format for scholarly data publishing that has received considerable uptake in the last few years. In contrast to the common Linked Data publishing practice, nanopublications work at the granular level of atomic information snippets and provide a consistent container format to attach provenance and metadata at this atomic level. While the nanopublications format is domain-independent, the datasets that have become available in this format are mostly from Life Science domains, including data about diseases, genes, proteins, drugs, biological pathways, and biotic interactions. More than 10 million such nanopublications have been published, which now form a valuable resource for studies on the domain level of the given Life Science domains as well as on the more technical levels of provenance modeling and heterogeneous Linked Data. We provide here an overview of this combined nanopublication dataset, show the results of some overarching analyses, and describe how it can be accessed and queried.
\end{abstract}


\section{Introduction}

Provenance has been identified as a crucial aspect to enable trust in Linked Data environments in general and eScience in particular \cite{moreau2011provenance}, and has led to the now widely adopted PROV ontology \cite{lebo2013prov}. However, there is still a lack of proven methods on how to represent and publish such provenance information for scientific data in a general, reliable, and agreed-upon manner. Nanopublications \cite{groth2010anatomy,mons2011value} have been proposed as a solution to this problem by providing a granular and principled way of publishing scientific (and other types of) data in a provenance-centric manner. Such a nanopublication consists of an atomic snippet of a formal statement (called ``assertion'') that comes with information about where this knowledge came from (the provenance of the assertion, i.e. how it was discovered) and with metadata about the nanopublication as a whole (the provenance of the nanopublication, i.e. how it was created; this part is called ``publication info''). All these three parts are represented as Linked Data (in RDF) and together they constitute a self-contained entity that we call a nanopublication. On the technical level, nanopublications are implemented with the help of named RDF graphs \cite{carroll2005named}, with one graph for each of assertion, provenance, and publication info, plus an additional head graph that holds everything together. Here, we present a growing dataset of such nanopublications, currently consisting of more than 10 million nanopublications containing diverse data (mostly from the Life Sciences so far) coming from a variety of contributors.

\begin{figure}[tb]
\begin{center}
\includegraphics[width=\columnwidth]{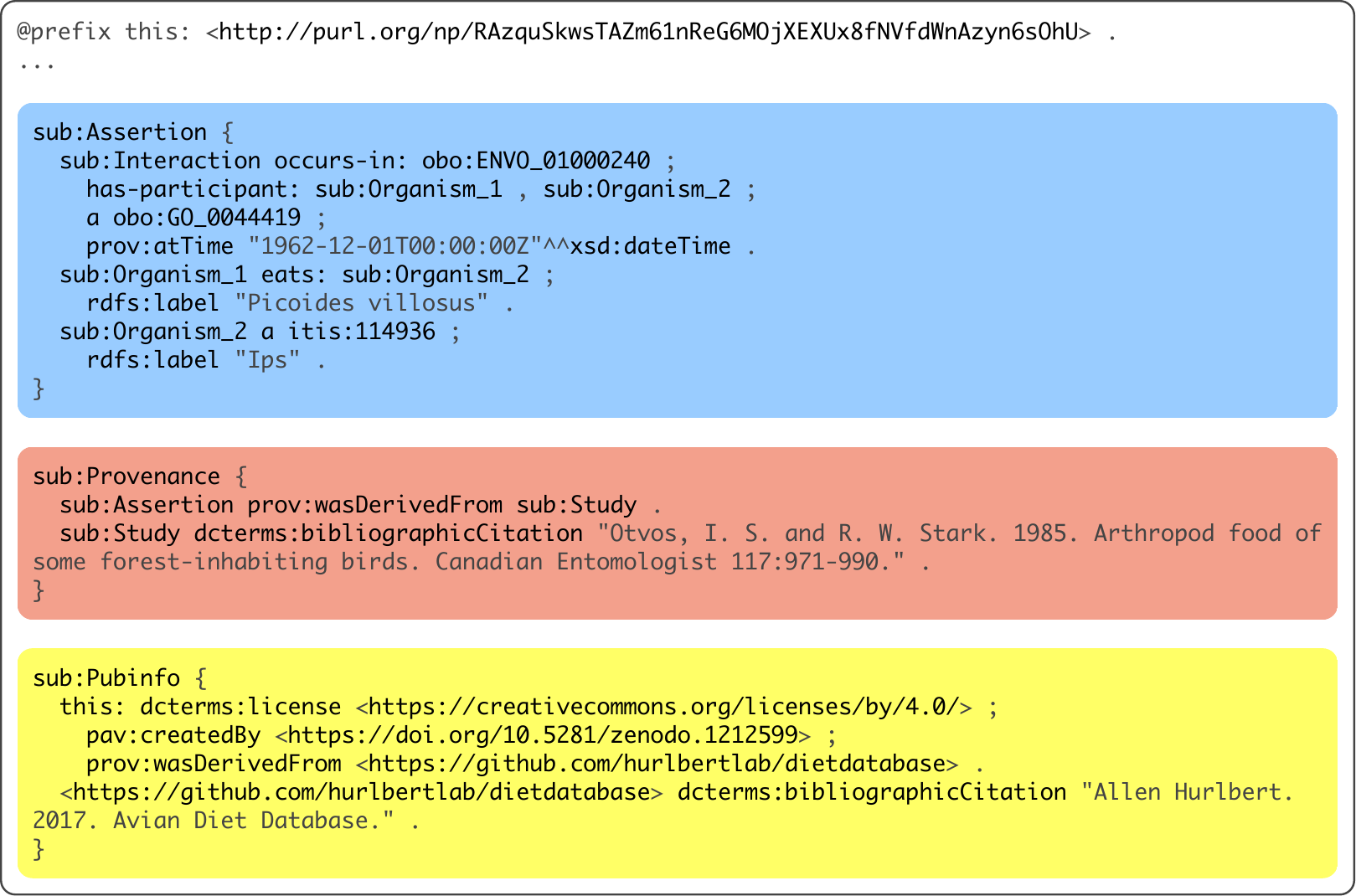}
\caption{\label{fig:nanopub-example} An example of a nanopublication on bird diets}
\end{center}
\end{figure}

Figure \ref{fig:nanopub-example} shows an example of a nanopublication from a recently added dataset on bird diets. It shows the three main named graphs (the head graph is not shown), and the prefix of the first line shows the nanopublication's identifier (not shown are all other prefixes). The assertion states that there is an interaction of type ``interspecies interaction between organisms'' (\texttt{obo:GO\_0044419}) with two participating organisms occurring in a place of ``conifer woodland'' (\texttt{obo:ENVO\_01000240}) on a given day in 1962. We are further told that the interaction was in fact that one of the organisms ate the other, where the eating organism is described with the species label ``Picoides villosus'' and the eaten organism is classified under the genus with a certain taxonomic serial number (\texttt{itis:114936}) labeled as ``Ips''. If we are wondering about how we happen to know about this particular ``interaction'', we can look at the provenance graph, where we read that this assertion was derived from a study reported in a paper published in 1985. While the provenance part tells us where the knowledge encoded in the assertion came from, the publication info part tells us more about how the triples that make up this nanopublication were created. We see that it was created by a thing identified with a DOI (\url{https://doi.org/10.5281/zenodo.1212599}, which happens to be a software tool), derived from a GitHub repository (\texttt{hurlbertlab/dietdatabase}) attributed to Allen Hurlbert in 2017. We see that the recorded interaction, the recording of the interaction, and the publication in the current form as a nanopublication happened in very different contexts spread over many decades. It thereby also demonstrates the importance of differentiating between them in an unambiguous and systematic manner, which is exactly the strength and purpose of the nanopublication format.

\section{Background}

In previous work, we addressed the problems of the reliability and persistence of such nanopublications and of principled and efficient versioning for evolving datasets. Nanopublications were from the start defined as being immutable, but only with our work on trusty URIs \cite{kuhn2014trusty,kuhn2015making} we could provide technical guarantees on immutability with the use of cryptographic hashes. Nanopublications are thereby identified by URIs that contain a hash value that is calculated on their entire content, whereby even minimal changes can be detected and users can formally verify nanopublication data against their identifiers, to ensure the data perfectly corresponds to what they were looking for. An example of such a trusty URI is shown in the first line of Figure \ref{fig:nanopub-example}.

In order to allow users to reliably retrieve nanopublications given their URIs, we built upon this work to develop a decentralized publishing network \cite{kuhn2015publishing,kuhn2016peerjcs}, currently consisting of fifteen server instances in nine countries\footnote{\url{http://purl.org/nanopub/monitor}}. We showed that nanopublication datasets can be efficiently and reliably archived and retrieved, without depending on the uptimes of individual servers. This server network is fully open, in the sense that everybody can publish nanopublications through it and even set up their own server to become a node in the network.

In order to account for the fact that we are often interested in entire datasets and not just individual data entries, and that such datasets change over time, we developed the concept of nanopublication indexes, which are themselves nanopublications. Such indexes point to other nanopublications, thereby defining sets of them of arbitrary size, which can also be used to define incremental dataset versions that reuse as much as possible from the previous versions \cite{kuhn2017reliable}. We could demonstrate that this kind of dataset representation and versioning is indeed efficient and effective, and that the overheads implied by nanopublications are offset by the benefits of referencing and using specific well-defined subsets.

On a more practical level, we developed a Java nanopublication library \cite{kuhn2015nanopub} and a command-line utility tool called npop\footnote{\url{https://github.com/tkuhn/npop}} to support the development of more high-level applications.

\section{Related Work}

Scientific data publishing has lately been a very popular topic and led to the proposal and adoption of the FAIR guiding principles for scientific data management and stewardship \cite{wilkinson2016fair}, mandating scientific data to be Findable, Accessible, Interoperable, and Reusable. Linked Data approaches directly tackle the interoperability challenge, and the Linked Open Data cloud \cite{bizer2009linked} has come to contain large amounts of scientifically relevant structured data. However, we are still lacking accepted technologies to directly publish evolving Linked Data resources in a reliable manner, and much of the existing Linked Open Data is in fact extracted with custom scripts from different heterogeneous data formats, e.g. via the Bio2RDF initiative \cite{dumontier2014bio2rdf}.

Apart from the nanopublication approach on which we focus in this paper, a number of other proposals address scientific data publishing. Solutions based on HTML for scholarly communication, such as RASH \cite{diiorio2015iswcpd} and Dokieli \cite{capadisli2017icwe}, start from scientific articles and annotate them with Linked Data via HTML, RDFa, and related technologies. Other approaches like Research Objects \cite{bechhofer2013future} embrace a broader variety of types of digital resources, such as data tables, metadata, source code, presentation slides, log files, and workflow definitions, and provide a linked format for bundling them in well-defined packages. As a further notable approach, micropublications \cite{clark2014biomedsem} combine elements of the approaches above, while putting an emphasis on the structure of scientific arguments and their relation to pieces of evidence.
Nanopublications differ from these approaches in their exclusive focus on Linked Data (thereby not directly covering things like source code or diagrams) and in their emphasis of provenance at the most fine-grained level.

\section{Data}

The data presented here is a wild collection of various datasets, consisting in total of 10\,803\,231 nanopublications. It contains new datasets extracted and modeled from existing resources, including nanopublication datasets on DrugBank (via Bio2RDF) and GloBI, and two smaller new datasets that were modeled as nanopublications from the start (``nano-born'') on polycystic kidney disease and monogenic rare diseases, respectively. Furthermore, the collection also includes datasets derived from OpenBEL and the Human Protein Atlas, which are already a few years old but have until now not been properly introduced, as well as syntactically improved versions (including minting of trusty URIs) of previously introduced datasets on GeneRIF/AIDA and neXtProt. Moreover, the collection also contains new versions of the previously announced nanopublication datasets on DisGeNET (4 versions), WikiPathways (21 versions), and LIDDI (2 versions), the latter of which is also nano-born. Lastly, there is a small number of loose nanopublications that are not part of any dataset. We will now briefly describe all of these datasets.

The \emph{DrugBank/Bio2RDF} dataset was extracted from Bio2RDF resources \cite{dumontier2014bio2rdf} originating from the DrugBank database \cite{wishart2007drugbank}, including drug--drug interactions, drug targets, and food interactions. These nanopublications were automatically extracted with a detailed meta-model to describe (in the publication info part) the processes, installations, versions, and codebases that produced each individual nanopublication\footnote{\url{https://github.com/tkuhn/bio2rdf2nanopub}}.

\emph{Global Biotic Interactions} (GloBI) is an initiative that aims to make existing species interaction datasets more easily accessible to help address the Eltonian shortfall in openly accessible biodiversity data. Instead of acting as a centralized data repository, GloBI provides tools to continuously discover, integrate and aggregate existing datasets across general purpose repositories like GitHub and Zenodo in addition to supporting various specialized infrastructures like iNaturalist\footnote{\url{https://inaturalist.org}}, Web of Life\footnote{\url{http://www.web-of-life.es/}}, and the Interaction Bank of the Natural History Museum, London \cite{baker2016nhm}. Resulting aggregate datasets can now be indexed, linked, transformed and made available as data archives, APIs and, more recently, as nanopublications covering a subset of the data. By providing tools to find and link datasets that describe how organisms interact (e.g., parasite-host, pollinator-plant, pathogen-host) with each other and across other biodiversity data platforms, GloBI helps to better integrate valuable knowledge acquired in recent and not-so recent past. Apart from the iNaturalist database mentioned above, the current nanopublication subset also includes datasets on turfgrass diseases\footnote{\url{https://github.com/globalbioticinteractions/aps-turfgrasses}}, bees\footnote{\url{https://github.com/globalbioticinteractions/Catalogue-of-Afrotropical-Bees}}, interactions of ocean species\footnote{\url{https://github.com/globalbioticinteractions/raymond}}, and the \emph{Avian Diet Database}\footnote{\url{https://github.com/hurlbertlab/dietdatabase}}.
We have already seen an example of the latter in Figure \ref{fig:nanopub-example}. The Avian Diet Database is a compilation of quantitative diet data for bird species of North America extracted from the primary literature. Individual records describe a trophic link between a bird species and a prey item, and the strength of that link based on the fraction of the diet by weight, number of items, or occurrence. In addition, each trophic link is described by contextual information about the time and place of the diet study, sample size, and original citation.

A small dataset on a \emph{meta-analysis of polycystic kidney disease expression profiles} was derived from sequenced male mouse RNA data \cite{malas2017meta}. The data is split into three phases of disease progression: early, moderate, and advanced.
The mice of the different phase groups were sacrificed at different weeks after gene disruption and had different cystic phenotypes (mild, moderate, and severe). The results of this meta-analysis were then represented in 1657 nanopublications.

Another small dataset on \emph{monogenic rare diseases and their genes} was recently published in the nanopublication format. Starting from diseases and their causative genes in the OMIM database \cite{amberger2014omim}, a subset of disease-gene associations were selected for monogenic (disease is caused by a mutation in one gene), rare (occuring in less than 1:2000) diseases. For each of them, the reference stating for the first time that this disease is causally associated with by this gene was manually curated. These associations are now exposed as nanopublications, following the DisGeNET model~(see \cite{queralt2016publishing} and below). The nanopublications represents the diseases with their OMIM identifiers, genes with their HGNC symbols (provided by Wikidata~\cite{Mietchen2015wikidata,Putman2017wikigenomes}), and the Ensembl gene identifier (which was added using BioMart \cite{kinsella2011ensembl}) for the gene, and the PubMed identifier for the provenance.

The \emph{OpenBEL} dataset consists of nanopublications converted from the Biological Expression Language (BEL) format from resources provided by the OpenBEL initiative\footnote{\url{http://openbel.org/}}. The conversion process maps BEL-specific vocabularies to existing accepted ontologies, as far as possible\footnote{\url{https://github.com/tkuhn/bel2nanopub}}.

The \emph{Human Protein Atlas} (HPA) \cite{uhlen2010towards} is a dataset that concentrates on the genome-wide analysis of human proteins. The nanopublications are a subset of the entire dataset, presently covering the immunohistochemistry results. They were built on the neXtProt tissue expression nanopublication model (see below) with some minor modifications. For example, the Neuroscience Information Framework (NIF) Standard Ontology is used in the assertion graph to specify the quality of the immunohistochemical staining.

The \emph{neXtProt} database is a high-quality corpus of data describing human proteins. Three categories of nanopublications had been derived and published from the a subset of the neXtProt data \cite{chichester2015converting}. They detail information on the tissue expression of proteins, protein posttranslational modifications (PTM), and single amino acid polymorphisms, respectively. For the tissue expression nanopublications, the assertion graph contains the protein, tissue, and quality of the expression result, whereas the provenance graph contains the method of detection and an assessment of the evidence.  The PTM nanopublication assertion delineates the specific modification in one specific protein isoform. The nanopublications for the amino acid polymorphisms describe the codon or codons variations that result in a protein change. Since their first publication, these neXtProt nanopublications have been syntactically improved by minting trusty URIs for them.

The \emph{GeneRIF/AIDA} dataset contains sentences in a controlled natural language that were generated from input of the GeneRIF dataset on gene functions \cite{kuhn2013broadening}. As with the neXtProt dataset, these nanopublications have been improved since their original publication by including trusty URIs.

Starting from version v2.1.0.0, \emph{DisGeNET} has released its linked dataset on human gene-disease associations (GDAs) also in the form of nanopublications \cite{queralt2016publishing}. That first nanopublication version consisted of 940\,034  nanopublications, representing the same number of scientific assertions for 381\,056 different GDAs with their detailed provenance, levels of evidence and publication information descriptions. Since then, three more versions have been released: v3.0.0.0, v4.0.0.0, and v5.0.0.0. The assertion part of these nanopublications contains the description for a specific single GDA, and the provenance graph includes provenance, evidence and attribution statements that were directly mapped from the VoID description of DisGeNET's RDF representation. With each DisGeNET database release, the data collected from the existing data sources are updated and new data sources are added. BeFree is an important data source, providing a growing dataset generated by text-mining millions of biomedical abstracts from Medline. In the last version v5.0.0.0, there are 1\,469\,541 nanopublications for 561\,119 GDAs between 17\,074 genes and 20\,370 diseases, disorders, traits, and clinical or abnormal human phenotypes.

\emph{WikiPathways} is a database for biological pathways, including metabolic, signaling and genetic pathways~\cite{slenter2017wikipathways}. Since 2016 WikiPathways publishes their monthly data releases also as in the nanopublication format, which has led to than twenty new versions since the initial release~\cite{kuhn2017reliable}. These nanopublications still describe only a subset of the knowledge in the database, focusing and exposing facts that are explicitly backed by literature, identified with PubMed identifiers. The assertions cover participation of genes, proteins, and metabolites in pathways, complexes, and interactions between these entities in pathways.

The \emph{LIDDI} dataset on Linked Drug-Drug Interactions, finally, was described in an earlier publication \cite{banda2015provenance}, but a new version has since been released. In this updated version, the drug mappings have been cleaned and the drug-drug-adverse event triples properly linked. Multiple research groups have started using the dataset and some of their suggestions for improvements have been incorporated. For an upcoming more substantial release, coverage of more drugs and more adverse events is planned.

All these datasets and their versions are defined by nanopublication indexes, which link to the respective sets of nanopublications. As nanopublication indexes are nanopublications themselves, they can be published via the same server network and form part of the nanopublication collection as regular data entries. Tables \ref{tab:indexes1} and \ref{tab:indexes2} at the end of this paper show all indexes that have been published until now, 129 in total, including some experimental unnamed indexes. \emph{Date} denotes the creation date of the index, and \emph{Sub} shows the number of sub-indexes. The WikiPathways and OpenBEL datasets make use of such sub-indexes to partition their data into several referenceable subsets. For example, index number 6 (first version of OpenBEL dataset) has indexes number 1 and 2 as sub-indexes and therefore contains the union of their nanopublications. Such subsets can also be defined post-hoc on and across existing datasets, for example index number 11 in Table \ref{tab:indexes1}, which includes nanopublications from the OpenBEL and neXtProt datasets. The column \emph{Size}, finally, shows the number of contained nanopublications (including sub-indexes).

\section{Analysis}

The unifying format of nanopublications now allows us to make an overarching analysis over all these heterogeneous datasets in a complete and uniform manner. The 10\,803\,231 nanopublications are in made up of 378\,654\,287 triples in total, 61\,184\,484 in the assertion graphs, 122\,229\,003 in the provenance graphs, and 136\,738\,995 in the publication info graphs. We see that these datasets are indeed ``provenance-centric'' with an average of 11.3 provenance triples per nanopublication. The average size of a nanopublication is 35.1 triples.

\begin{table*}[tb]
\begin{center}
\caption{Creators and authors of nanopublications}
\label{tab:creators}
\begin{tabular}{r@{~~}|@{~~}r@{~~}r@{~~}|@{~~}l}
type & total & unique & example (with max. frequency) \\
\hline
ORCID & 40\,964\,679 & 26 & \url{http://orcid.org/0000-0003-0169-8159} \\
Literal string & 6\,534\,917 & 3 & {\tt"CALIPHO project"} \\
Tool URI & 79\,617 & 4 & \url{https://doi.org/10.5281/zenodo.1212599} \\
Google Scholar URI & 3 & 3 & \url{https://scholar.google.it/citations?user=9aI21r8AAAAJ&hl=en} \\
ResearcherID & 3 & 3 & \url{http://www.researcherid.com/rid/B-6035-2012} \\
Other URI & 16 & 2 & \url{http://sorry.vse.cz/~xhudj19} \\
\hline
Total & 47\,579\,235 & 41 & \\
\end{tabular}
\end{center}
\end{table*}

Next, we wanted to get an idea of who created these nanopublications. For that, we ran a SPARQL query to find all identifiers that were listed as creators or authors of a nanopublication in its publication info part. Specifically, we considered the predicates \texttt{dct:creator}, \texttt{dce:creator}, \texttt{pav:createdBy}, \texttt{pav:authoredBy}, and \texttt{prov:wasAttributedTo}. The summary of the result is shown in Table \ref{tab:creators}.
In total there are more than 47 million such creator mentions, i.e. an average of 4.4 creators per nanopublication. This set is however dominated by a much smaller number of ``power-users'' with just 41 unique identifiers. By far the most widely adopted identification scheme is ORCID (86\% of all mentions), followed by a somewhat irregular use of literal strings (14\%), and URIs identifying software tools (0.17\%).

\begin{figure}[tb]
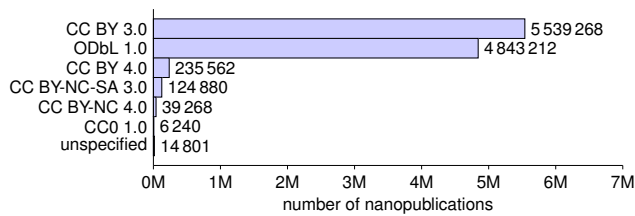

\begin{center}\scalebox{0.65}{
\begin{bchart}[max=7,step=1,unit=M,scale=0.8,width=12cm]
\bcbar[label=CC BY 3.0,value=5\,539\,268]{5.539268}
\bcbar[label=ODbL 1.0,value=4\,843\,212]{4.843212}
\bcbar[label=CC BY 4.0,value=235\,562]{0.235562}
\bcbar[label=CC BY-NC-SA 3.0,value=124\,880]{0.124880}
\bcbar[label=CC BY-NC 4.0,value=39\,268]{0.039268}
\bcbar[label=CC0 1.0,value=6\,240]{0.006240}
\bcbar[label=unspecified,value=14\,801]{0.014801}
\bcxlabel{number of nanopublications}
\end{bchart}
}\end{center}
\caption{\label{fig:licenses} License distribution}
\end{figure}

Next, we can turn to the topic of licensing. Figure \ref{fig:licenses} shows the distribution of licenses as reported in the nanopublications, considering \texttt{dct:license} and \texttt{dct:rights}. Apart from DisGeNET using the Open Database License (ODbL), Creative Commons (CC) licenses dominate the dataset, mostly in its CC BY flavor in version 3.0 and (to a much lesser extent) 4.0. This effect is mainly driven by a few large datasets that use CC BY 3.0, namely WikiPathways, neXtProt, and the Human Protein Atlas. (WikiPathways has just switched to CC0 1.0 for future versions.)



Next, we can look a bit more deeply to find out what these nanopublications are really about. We can do this by counting the used namespaces (i.e. the shared first part of URIs that group identifiers into vocabularies and collections) in the four different graphs (head, assertion, provenance, and publication info) and the three different triple positions (subject, predicate, and object).
Table \ref{tab:namespaces} shows the most relevant namespaces in the nanopublications datasets, according to their frequency of occurrence in the subject, predicate and object position for all head, assertion, provenance, and publication info graphs. The shown percentages denote the ratio of nanopublications where the given namespace appears at least once in the given position.

\begin{table*}[t]
 \caption{Namespaces and their frequency in the different positions in nanopublications. The full URIs for the shown prefixes are shown at the bottom.}
 \label{tab:namespaces}
\setlength{\tabcolsep}{8pt}
\begin{center}
 \small
 \begin{tabular}{l|lr|lr|lr|}
\textbf{Graph} & \textbf{subj} & & \textbf{pred} &  & \textbf{obj} & \\
 \hline
\rowcolor{hcolor} head      & {nxpt} & 37.27\%  & {rdf} & 100.00\% & {np} & 100.00\% \\
\rowcolor{hcolor}           & {disgen} & 36.13\% & {np} & 100.00\% &  {nxpt} & 37.27\% \\
\rowcolor{hcolor}           & {panp} & 11.61\% & {rdfs} & 11.61\% & {disgen} & 22.53\% \\
\rowcolor{hcolor}           & {disgent} & 8.70\% & & & {disgent} & 8.70\% \\
\rowcolor{hcolor}           & {liddi-r} & 1.82\% & & & & \\
\rowcolor{hcolor}           & {klab} & 1.44\% & &  &  & \\
\rowcolor{hcolor}           & {nnp} & 1.41\% & &  & & \\
\rowcolor{hcolor}           & {bel2-r} & 1.16\% &  &  & & \\
\rowcolor{hcolor}           & {dbv} & 0.25\%  &  &  & & \\
\rowcolor{hcolor}           & {dbr} & 0.11\% & &  & & \\
 \hline
\rowcolor{acolor} assertion & {gda-r} & 36.13\% & {rdf} & 69.61\% & {sio} & 44.87\% \\
\rowcolor{acolor}           & {ncbi} & 31.27\% & {sio} & 52.43\% &  {ncbi} & 44.85\% \\
\rowcolor{acolor}           & {umls} & 31.22\% & {obo} & 36.24\% & {umls} & 44.83\% \\
\rowcolor{acolor}           & {nxpts} & 30.24\% & {oborel} & 30.24\% & {evs} & 31.27\% \\
\rowcolor{acolor}           & {pa} & 11.61\% & {obox} & 13.50\% & {caloha} & 30.24\% \\
\rowcolor{acolor}           & {gda} & 8.70\% & {nif} & 11.61\% & {pa} & 11.61\% \\
\rowcolor{acolor}           & {nxpt} & 7.02\% & {rdfs} & 11.15\% & {caloha-l} & 11.61\% \\
\rowcolor{acolor}           & {liddi-r} & 1.82\% & {dc} & 2.44\% & {nxpt} & 7.06\% \\
\rowcolor{acolor}           & {liddi} & 1.81\% & {dbv} & 2.17\% & {nxpts} & 7.02\% \\
\rowcolor{acolor}           & {nxptt} & 1.02\% & {npx} & 1.58\% & {rsp} & 7.02\% \\
 \hline
\rowcolor{pcolor} provenance & {nxpt} & 37.27\% &{prov} & 98.95\% & {eco} & 45.97\% \\
\rowcolor{pcolor}           & {nxptq} & 36.25\% & {rdf} & 95.66\% &  {pubmed-id} & 44.09\% \\
\rowcolor{pcolor}           & {disgen} & 22.53\% & {wi} & 93.71\% & {nxpt} & 37.27\% \\
\rowcolor{pcolor}           & {disgen-void5} & 13.60\% & {rdfs} & 83.25\% & {nxptq} & 37.27\% \\
\rowcolor{pcolor}           & {disgen-void4} & 13.10\% & {dc} & 48.38\% & {obox} & 36.13\% \\
\rowcolor{pcolor}           & {panp} & 11.61\% & {pprov} & 47.86\% & {efo} & 21.49\% \\
\rowcolor{pcolor}           & {disgen-void3} & 9.43\% & {sio} & 47.55\% & {bao} & 14.76\% \\
\rowcolor{pcolor}           & {bao} & 8.75\% & {pav} & 37.36\% & {disgen-void5} & 13.60\% \\
\rowcolor{pcolor}           & {disgen-void2} & 8.70\% & {owl} & 8.75\% & {disgen-void4} & 13.10\% \\
\rowcolor{pcolor}           & {disgent} & 8.70\% & {pav2} & 8.70\% & {pas} & 11.61\% \\
 \hline
\rowcolor{icolor} pubinfo   & {nxpt} & 37.27\% & {dc} & 100.00\% &{orcid} & 98.67\% \\
\rowcolor{icolor}           & {disgen} & 36.13\% & {pprov} & 93.71\% & {cc} & 50.69\% \\
\rowcolor{icolor}           & {disgen-void5} & 13.60\% & {swan} & 49.61\% & {http://} & 48.93\% \\
\rowcolor{icolor}           & {disgen-void4} & 13.10\% & {prov} & 40.81\% & {sio} & 44.83\% \\
\rowcolor{icolor}           & {panp} & 11.61\% & {pav} & 39.28\% & {odbl} & 44.83\% \\
\rowcolor{icolor}           & {disgen-void3} & 9.43\% & {rdf} & 38.02\% & {nxpt} & 37.27\% \\
\rowcolor{icolor}           & {disgen-void2} & 8.70\% & {pav2} & 8.70\% & {eco} & 37.27\% \\
\rowcolor{icolor}           & {disgent} & 8.70\% & {rdfs} & 1.54\% & {disgen-void5} & 13.60\% \\
\rowcolor{icolor}           & {liddi-r} & 1.82\% & {npx} & 0.56\% & {disgen-void4} & 13.10\% \\
\rowcolor{icolor}           & {klab} & 1.44\% & {dce} & 0.13\% & {rid} & 11.61\% \\

\hline
\end{tabular}
\end{center}%
\begin{center}
\scriptsize
\setlength{\tabcolsep}{1pt}
\begin{tabular}{lp{7.5cm}}
\textbf{Prefix} & \textbf{Namespace} \\
 \hline
bel2-r & \url{http://www.tkuhn.ch/bel2nanopub/} \\
bao & \url{http://www.bioassayontology.org/bao#} \\
caloha & \url{ftp://ftp.nextprot.org/pub/current_release/controlled_vocabularies/caloha.obo#} \\
caloha-l & \url{http://purl.obolibrary.org/obo/caloha.obo#} \\
cc & \url{http://creativecommons.org/licenses/by/} \\
dbv & \url{http://bio2rdf.org/drugbank_vocabulary} \\
dbr & \url{http://bio2rdf.org/drugbank:} \\
dc & \url{http://purl.org/dc/terms/} \\
dce & \url{http://purl.org/dc/elements/1.1/} \\
disgen & \url{http://rdf.disgenet.org/resource/nanopub/} \\
disgen-void2 & \url{http://rdf.disgenet.org/v2.1.0/void.ttl#} \\
disgen-void3 & \url{http://rdf.disgenet.org/v3.0.0/void/} \\
disgen-void4 & \url{http://rdf.disgenet.org/v4.0.0/void/} \\
disgen-void5 & \url{http://rdf.disgenet.org/v5.0.0/void/} \\
disgent & \url{http://rdf.disgenet.org/nanopublications.trig#} \\
eco & \url{http://purl.obolibrary.org/obo/eco.owl#} \\
efo & \url{http://www.ebi.ac.uk/efo/} \\
evs & \url{http://ncicb.nci.nih.gov/xml/owl/EVS/Thesaurus.owl#} \\
gda & \url{http://rdf.disgenet.org/gene-disease-association.ttl#} \\
gda-r & \url{http://rdf.disgenet.org/resource/gda/} \\
obo & \url{http://purl.obolibrary.org/obo/#} \\
oborel & \url{http://purl.org/obo/owl/OBO_REL#} \\
obox & \url{http://purl.obolibrary.org/obo/} \\
klab & \url{http://krauthammerlab.med.yale.edu/nanopub/} \\
liddi & \url{http://liddi.stanford.edu/LIDDI} \\
liddi-r & \url{http://liddi.stanford.edu/LIDDI_resource:} \\
ncbi & \url{http://identifiers.org/ncbigene/} \\
\end{tabular}
\quad
\begin{tabular}{lp{7.5cm}}
\textbf{Prefix} & \textbf{Namespace} \\
 \hline
nif & \url{http://ontology.neuinfo.org/NIF/Backend/NIF-Quality.owl#} \\
np & \url{http://www.nanopub.org/nschema#} \\
nnp & \url{http://purl.org/np/} \\
npx & \url{http://purl.org/nanopub/x/} \\
nxpt & \url{http://www.nextprot.org/nanopubs#} \\
nxptq & \url{http://www.nextprot.org/help/quality_criteria/} \\
nxpts & \url{http://www.nextprot.org/db/search#} \\
nxptt & \url{http://www.nextprot.org/db/term/} \\
odbl & \url{http://opendatacommons.org/licenses/odbl/} \\
orcid & \url{http://orcid.org/} \\
owl & \url{http://www.w3.org/2002/07/owl#} \\
pa & \url{http://www.proteinatlas.org/} \\
pas & \url{http://www.proteinatlas.org/search/} \\
pav & \url{http://purl.org/pav/} \\
pav2 & \url{http://purl.org/pav/2.0/} \\
panp & \url{http://www.proteinatlas.org/about/nanopubs/} \\
prov & \url{http://www.w3.org/ns/prov#} \\
pprov & \url{http://purl.org/net/provenance/ns#} \\
pubmed-id & \url{http://identifiers.org/pubmed/} \\
rid & \url{http://www.researcherid.com/rid/} \\
rdf & \url{http://www.w3.org/1999/02/22-rdf-syntax-ns#} \\
rdfs & \url{http://www.w3.org/2000/01/rdf-schema#} \\
rsp & \url{http://sadiframework.org/ontologies/GMOD/RangedSequencePosition.owl#} \\
sio & \url{http://semanticscience.org/resource/} \\
umls & \url{http://linkedlifedata.com/resource/umls/id/} \\
swan & \url{http://swan.mindinformatics.org/ontologies/1.2/pav/} \\
wi & \url{http://purl.org/ontology/wi/core#} \\
 \end{tabular}
\end{center}
\end{table*}

In the head graph, we always find predicates in the \texttt{rdf} and \texttt{np} namespaces, which is unsurprising since otherwise these would not be valid nanopublications. Subjects and objects in the head mostly denote the nanopublication itself and its graphs, respectively, and thereby tend to come from dataset-specific namespaces. 
In the assertion graph, \texttt{rdf} is the most frequent namespace in predicate position, mostly due to instantiations (\texttt{rdf:type}), but otherwise this category is --- unsurprisingly --- dominated by domain vocabularies like \texttt{sio}, \texttt{obo}, \texttt{oborel}, and \texttt{obox}. This is even more the case for the object position with \texttt{sio}, \texttt{ncbi} and \texttt{umls} being the most frequent namespaces, whereas the subject position is a mix of domain-specific and dataset-specific vocabularies.
In the provenance graphs, we see the classical provenance and metadata vocabularies in predicate position, including \texttt{prov}, \texttt{rdf}, \texttt{rdfs}, \texttt{dc}, \texttt{pav}, and \texttt{owl}. It is notable that \texttt{prov} reaches near-universal acceptance by occurring in 98.95\% of all nanopublications in that position. The Weighted Interests Vocabulary (\texttt{wi}) is also surprisingly popular, appearing in more than 93\% of all nanopublications. This provenance information is supposed to be attached to the assertion graph, which explains the domain-specific namespaces in subject position. In object position we see a mix of domain-specific and dataset-specific namespaces, which can be explained by provenance pointing to external domain entities as well as internal activities or objects.
The publication info graph is about the provenance of the nanopublication as a whole, and it is therefore not surprising to find domain-specific namespaces in subject position, and the classical provenance ontologies again in predicate positions, like \texttt{prov} and \texttt{pav}. DC Terms (\texttt{dc}) even achieves perfect adoption by appearing in 100\% of nanopublications in a predicate of this graph. In the object position, we find near-universal adoption of \texttt{orcid} at 98.67\%. Creative Commons license URIs (\texttt{cc}) and plain domain names of websites (like \url{http://nextprot.org}, which are summarized by the prefix \texttt{http://}) are popular as well.
This kind of analysis provides us with interesting insights into the content of such a large number of nanopublications, but we also have to be aware that these numbers are often driven by a few large datasets.

\begin{figure}[tb]
\begin{center}
\includegraphics[width=0.9\columnwidth]{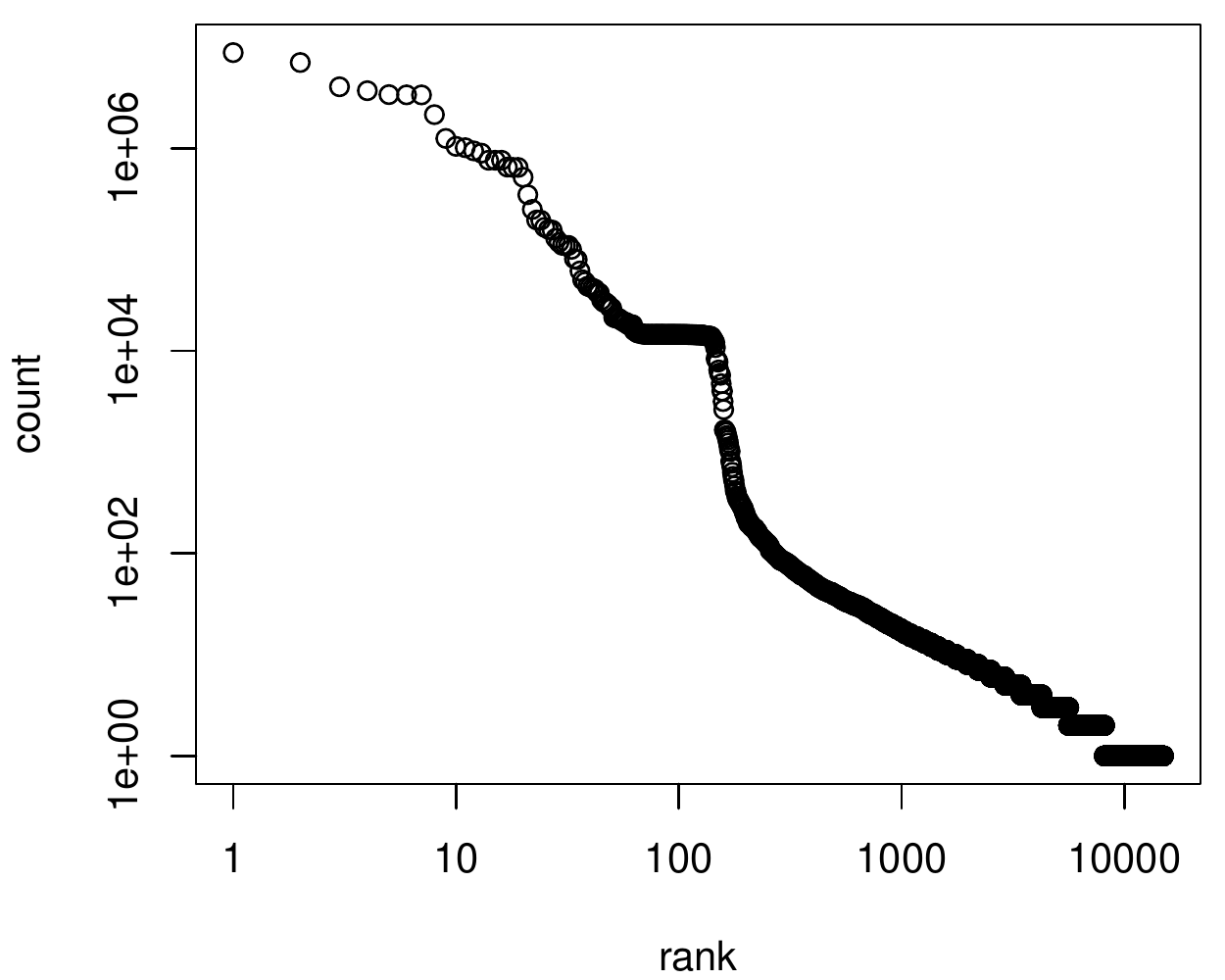}
\end{center}
\caption{\label{fig:typedist} Frequency distribution of types}
\end{figure}

Finally, we can investigate a bit more the variety of content found in these nanopublications. For this we filtered all triples from the assertion graphs that have \texttt{rdf:type} as predicate, thereby assigning an individual to a type. Overall, we found 50\,384\,007 such individual--type assignments, involving 14\,941 unique types. The most frequent type, \texttt{eco:ECO\_0000218} representing ``manual assertion'', occurs 8\,828\,067 times, whereas on the other end of the scale many types such as \texttt{https://www.inaturalist.org/taxa/104422} (standing for Spotted Spreadwing, a species of damselflies) appear just once. Figure \ref{fig:typedist} visualizes the frequency distribution of these types, showing the continuous occurrence of classes along the whole size spectrum, which can be seen as an indication that the overall dataset is indeed varied and broad. The plateau in the middle is caused by the Human Protein Atlas via its reporting of the occurrence of proteins in a larger number of different tissues, including the cases where no information about the occurrence is available. Thereby, each of the tissues produces a group of nanopublications of a size that matches the number of covered proteins (around 14\,500), thereby producing this plateau effect in the diagram.

\section{Availability}

All nanopublications of the data we presented here can be directly retrieved from the decentralized server network. Most of the nanopublication URIs directly resolve to a node of the network, and in case this does not work they can all be found on various servers in the network via a well-defined procedure \cite{kuhn2015publishing}. To retrieve individual nanopublications or entire datasets in a more efficient way, the nanopublication library or its command-line interface can be used \cite{kuhn2015nanopub}. A nanopublication dump of the complete nanopublication content of the server network is furthermore made available on Zenodo\footnote{\url{https://doi.org/10.5281/zenodo.1213293}}.

In order to facilitate easier and more powerful access to nanopublications, we also developed a Linked Data API to access the full set of nanopublications available on the network. This API is powered by a \texttt{grlc} server \cite{meronyo2017grlc} at the front and a GraphDB triple store instance with a SPARQL endpoint at the back.\footnote{The API is available at \url{http://purl.org/nanopub/api}} This API offers a standard entry point to the data in the nanopublications network, which any Linked Data client can consume via HTTP without specific knowledge of SPARQL or RDF.\footnote{The underlying parametrized queries can be found at \url{https://github.com/peta-pico/nanopub-api/}} Concretely, our API provides the following methods:
\begin{itemize}
\item \textbf{find\_latest\_nanopubs\_with\_pattern}. Find all nanopublications for a given triple pattern, sorted by recency
\item \textbf{find\_nanopubs\_with\_pattern}. Find all nanopublications for a given triple pattern, in undefined order (faster than the method above)
\item \textbf{find\_latest\_nanopubs\_with\_uri}. Find all nanopublications that contain the given URI, sorted by recency
\item \textbf{find\_nanopubs\_with\_uri}. Find all nanopublications that contain the given URI, in undefined order (faster than the method above)
\item \textbf{get\_all\_indexes}. Get all nanopublication indexes (excluding incomplete ones); this gives a result similar to Tables \ref{tab:indexes1} and \ref{tab:indexes2}
\item \textbf{get\_index\_elements}. Get all nanopublications that the given index directly contains as elements
\item \textbf{get\_nanopub}. Get nanopublication by URI identifier (alternatively, this can also be achieved by directly calling the servers of the nanopublication network)
\end{itemize}
The GraphDB instance automatically loads nanopublications published to the network, normally within a few minutes or less.

Lastly, we also published the source code used for the shown analyses and processes on GitHub\footnote{\url{https://github.com/tkuhn/nanoresource}} and archived it on Zenodo\footnote{\url{https://doi.org/10.5281/zenodo.1213690}}.

\section{Discussion}

The nanopublication approach can be seen as an instance of ``containerization'' analogous to the industrial containerization for efficient transport of goods in standardized, physical containers (and to the more recent containerization of software with solutions like Docker). Efficiency is improved with a standardized form and size of the containers, which allows for large-scale automatic and reliable processing of the content (data and physical goods, respectively). The diverse dataset presented here illustrates the recent uptake and expected benefits of this nanopublication-based containerization approach for scientific data publishing.

The current collection of nanopublications can be a valuable resource on the domain level, in particular for biomedical studies where reliability, reproducibility, and trust are important. Our dataset provides good coverage on the biomedical domains of genes, proteins, diseases, biological pathways, and drugs.

Furthermore, our data may also prove to be valuable to study the modeling and processes related to provenance in general and the PROV ontology in particular, with more than 122 million provenance triples, and with more than 10 million nanopublications using the PROV ontology.

Lastly, the dataset can also be useful on a lower technical level, as a dataset with a very large number of graphs. The LIDDI dataset has already been used for benchmarking named graph handling \cite{fernandez2018hdtq}, highlighting LIDDI's ``extremely large number of graphs, 392,340''. The combined nanopublication dataset presented here is in this sense more than 100 times more extreme.
LOD Laundromat \cite{beek2014lod}, as another point of comparison, contains about 100 times more triples (currently more than 38 billion), but only about 1.5\% the number of named graphs of our nanopublication dataset (658\,045 documents with a named graph each in LOD Laundromat, compared to more than 43 million named graphs in our nanopublication dataset).

As future work, we have concrete plans to attract further datasets, including the entire Bio2RDF database and larger subsets of resources such as GloBI. We are also working to improve the API and to establish a whole ecosystem of decentralized services feeding from the data backbone of the nanopublication network. In order make the execution of API calls more efficient and scalable, we will also look into the techniques of HDT-Quads \cite{fernandez2018hdtq} and the quad version of Triple Pattern Fragments \cite{verborgh2016triple}, as soon as these technologies become stable enough for this kind of application.

\begin{table*}[tb]
\centering
\caption{Overview of nanopublication indexes (continued in Table \ref{tab:indexes2})}
\label{tab:indexes1}
\small
\begin{tabular}{rllrr}
\# & Index Title (some abbreviated) & Date & Sub & Size \\
\hline
1 & \href{http://np.inn.ac/RAR7OfS-AqG9_XogObQZpWq6LaBsV95jeseJtscuwpwJo}{OpenBEL's Small Corpus 1.0} & 2015-03-02 & & 2033 \\
2 & \href{http://np.inn.ac/RAsmhDKpyhTORhvWOZZL_cfSVZUqLYJFl_AqF_xPN34Qw}{OpenBEL's Large Corpus 1.0} & 2015-03-02 & & 48674 \\
3 & \href{http://np.inn.ac/RAtF0ivB9B8cb-u3K_zElgmRBxiDwfym1yVBRY6VAyWvE}{OpenBEL's Small Corpus 20131211} & 2015-03-02 & & 1288 \\
4 & \href{http://np.inn.ac/RAdw0S5f06-Ed2uXNISU-wbXmqvQ9-6hxd4fqIslT38Wg}{OpenBEL's Large Corpus 20131211} & 2015-03-02 & & 72885 \\
5 & \href{http://np.inn.ac/RAY_lQruuagCYtAcKAPptkY7EpITwZeUilGHsWGm9ZWNI}{AIDA GeneRIF} & 2015-03-04 & & 156026 \\
6 & \href{http://np.inn.ac/RACy0I4f_wr62Ol7BhnD5EkJU6Glf-wp0oPbDbyve7P6o}{OpenBEL's Small and Large Corpus 1.0} & 2015-03-04 &2 & 50707 \\
7 & \href{http://np.inn.ac/RAR5dwELYLKGSfrOclnWhjOj-2nGZN_8BW1JjxwFZINHw}{OpenBEL's Small and Large Corpus 20131211} & 2015-03-04 &2 & 74173 \\
8 & \href{http://np.inn.ac/RAXy332hxqHPKpmvPc-wqJA7kgWiWa-QA0DIpr29LIG0Q}{DisGeNET v2.1.0.0} & 2015-03-04 & & 940034 \\
9 & \href{http://np.inn.ac/RAXFlG04YMi1A5su7oF6emA8mSp6HwyS3mFTVYreDeZRg}{neXtProt protein data (preliminary)} & 2015-03-09 & & 4025981 \\
10 & \href{http://np.inn.ac/RAXsXUhY8iDbfDdY6sm64hRFPr7eAwYXRlSsqQAz1LE14}{\emph{(unnamed)}} & 2015-04-09 & & 3 \\
11 & \href{http://np.inn.ac/RA6jrrPL2NxxFWlo6HFWas1ufp0OdZzS_XKwQDXpJg3CY}{Data about CDKN2A from BEL2nanopub \& neXtProt} & 2015-04-14 & & 5 \\
12 & \href{http://np.inn.ac/RA7SuQ0e661LJdKpt5EOS2DKykf1ht9LFmNaZtFSDMrXg}{Linked Drug-Drug Interactions (LIDDI) dataset v1.01} & 2015-07-17 & & 98085 \\
13 & \href{http://np.inn.ac/RAFa_x4h0ng_NXtof35Ie9pQVsAY69Ab3ZQMir2NP8vGc}{\emph{(unnamed)}} & 2015-08-18 & & 3 \\
14 & \href{http://rdf.disgenet.org/nanopub-server/RAVEKRW0m6Ly_PjmhcxCZMR5fYIlzzqjOWt1CgcwD_77c}{DisGeNET v3.0.0.0} & 2015-11-15 & & 1018735 \\
15 & \href{http://np.inn.ac/RAnAgWoQuJYRDoC02iZdnPP8CHyDNHTY_HKVCj5BqwdxM}{\emph{(unnamed)}} & 2016-01-09 & & 2 \\
16 & \href{http://np.inn.ac/RAxM_p9YP3oZrQZQm_pDg_tVpUKg15UPy_VY90fXiF-NQ}{Regional goverments of Czech Republic} & 2016-01-09 & & 13 \\
17 & \href{http://rdf.disgenet.org/nanopub-server/RAyRv500NbzzqmkouEraxBX-Op9Hg4Wps0HrQkNFyeNLQ}{DisGeNET v4.0.0.0 (Nanopub Index)} & 2016-05-13 & & 1414902 \\
18 & \href{http://np.inn.ac/RAhoIw-XmyVv7DfMKyni67OHBFRqPvHAodr_iRgJxhtxY}{Rett Syndrome data from DisGeNET v3.0.0.0} & 2016-11-01 & & 1284 \\
19 & \href{http://np.inn.ac/RAwMqPOC0ZdZquSGih3aNsUuwGAGGtaCR3L4r5IRkK5x0}{Rett Syndrome data from DisGeNET v4.0.0} & 2016-11-02 & & 968 \\
20 & \href{http://np.inn.ac/RAwepzy1rtZCwp_ABboUHReraz95kc4-EM-RP6V4RkKzc}{\emph{(unnamed)}} & 2016-11-02 & & 1642 \\
21 & \href{http://np.inn.ac/RAK92fVJCMpJ3Ie3jMXiehtKnFNzvGd3MQnXOj55CgML8}{\emph{(unnamed)}} & 2016-11-02 & & 2566 \\
22 & \href{http://np.inn.ac/RAtyhkWi1DEMnlombfdWvO4zqi2OgCoNz3MRy-YucWH2Q}{\emph{(unnamed)}} & 2016-11-02 & & 692 \\
23 & \href{http://np.inn.ac/RA0ImMGdhttIJxDJsPw7HiRLZiqpIIhRhzw0Y6r8LvmKg}{\emph{(unnamed)}} & 2016-11-02 & & 233 \\
24 & \href{http://np.inn.ac/RANwPKrLBWo35dliB-75p4rnTt393KJK8LVZDiRN3mUZA}{Human Protein Atlas data} & 2016-11-02 & & 1254468 \\
25 & \href{http://np.inn.ac/RAV0-Gv-hPd05NWFQETchbhflTe5M44PolFlTkzwMWed4}{My example dataset} & 2016-11-28 & & 3 \\
26 & \href{http://purl.org/np/RA5rwWe7m-iWz8TF1kHIDCs9kLjwqIYeY3ZwLRQ5K0NZY}{\emph{(unnamed)}} & 2017-05-04 & & 1512 \\
27 & \href{http://purl.org/np/RADYX-ia_TZYAw_eZD0-2oGGA7gnMxOnVj-Gh8wdJgAzI}{DisGeNET v2.1.0.0, incremental dataset} & 2017-05-09 & & 940034 \\
28 & \href{http://purl.org/np/RAufQaKzv1pZlMhZo2eBuZtx9vuugLBJsrs4ZkvR53xzw}{DisGeNET v3.0.0.0, incremental dataset} & 2017-05-09 & & 1018735 \\
29 & \href{http://purl.org/np/RAu0PUrg-M8HxkOiYRXkTg7r9fgOIzFZNINj8q7ywNrdM}{DisGeNET v4.0.0.0, incremental dataset} & 2017-05-09 & & 1414902 \\
30 & \href{http://purl.org/np/RAiIF_QH1NK7I9AuGtPzCj_mP-jWv2beeZUTkEmk9PoW8}{complexes extracted from WikiPathways version 20160610} & 2017-05-11 & & 41 \\
31 & \href{http://purl.org/np/RAuBeU5Z_RW4snKOczFRB2mttOZ1ZzdUk8v4cUBkv6Tac}{complexes extracted from WikiPathways version 20160710} & 2017-05-11 & & 37 \\
32 & \href{http://purl.org/np/RAPABbweeUUbR4pcP8uCxwxLsSi_IJTiYp5dhTVjaSMkY}{complexes extracted from WikiPathways version 20160810} & 2017-05-11 & & 37 \\
33 & \href{http://purl.org/np/RAdGzmpzlOKGrFt4a-es2ncleen1al4AbYuA8tyr7Uk5A}{complexes extracted from WikiPathways version 20160910} & 2017-05-11 & & 37 \\
34 & \href{http://purl.org/np/RAVvPN0SI9dkGEO_kU-AeqI-llRl1gv7lOagAXKNAm7a8}{complexes extracted from WikiPathways version 20161010} & 2017-05-11 & & 37 \\
35 & \href{http://purl.org/np/RAp1UKyGTUgz-oxTJazoR7xUckYX_uwVd5JOoYVi60DKQ}{complexes extracted from WikiPathways version 20161110} & 2017-05-11 & & 36 \\
36 & \href{http://purl.org/np/RA7ursz5NEhudOkQxBgcaV8sw97RLna9FO7isyrypjW-g}{complexes extracted from WikiPathways version 20161210} & 2017-05-11 & & 37 \\
37 & \href{http://purl.org/np/RAN4lt0A-fANM4g1NuHni4J7xs1IlBaXYKVNGVoKBGOZo}{complexes extracted from WikiPathways version 20170210} & 2017-05-11 & & 42 \\
38 & \href{http://purl.org/np/RACwicgJPGLo5kW8r-4OE4PWpdXj1MEkSrAYmM1tm_kRU}{complexes extracted from WikiPathways version 20170310} & 2017-05-11 & & 42 \\
39 & \href{http://purl.org/np/RAND_wLT2kB7y5kU7XhM1_1v0FVuQsxlW2oHVUNODoYXM}{complexes extracted from WikiPathways version 20170410} & 2017-05-11 & & 42 \\
40 & \href{http://purl.org/np/RA1gsyHUYrQV3yOgRQ1XrHVWrrF41GQV1h6BePCfm-7eg}{complexes extracted from WikiPathways version 20170510} & 2017-05-11 & & 42 \\
41 & \href{http://purl.org/np/RA3-lOrJmxp2tYS3fa4t5dD09NcmcVbPeCV-suchaJFRA}{interactions extracted from WikiPathways version 20160610} & 2017-05-11 & & 8977 \\
42 & \href{http://purl.org/np/RAB4s_Mswg579pYGflUa1Gu_84hMmByKY8nn013npwBl0}{interactions extracted from WikiPathways version 20160710} & 2017-05-11 & & 10136 \\
43 & \href{http://purl.org/np/RAOHFPqgB7grnOPYfhXfcZEsY7dHHjfHQGzrL2Mp6aUtQ}{interactions extracted from WikiPathways version 20160810} & 2017-05-11 & & 10086 \\
44 & \href{http://purl.org/np/RA6Sbqku5f5D4rhHF9Sc-XVnCNDsk7CTrt24lnScnkHoA}{interactions extracted from WikiPathways version 20160910} & 2017-05-11 & & 10087 \\
45 & \href{http://purl.org/np/RA8nws6QKUl6NCEqOSBnV3VNdGoXTjDi2rSgFcciQotW4}{interactions extracted from WikiPathways version 20161010} & 2017-05-11 & & 10090 \\
46 & \href{http://purl.org/np/RAYuzoANh721DxMebtf-AZMyW3vIxBvZh22if0Pc8hNo0}{interactions extracted from WikiPathways version 20161110} & 2017-05-11 & & 10092 \\
47 & \href{http://purl.org/np/RABXul9kAvYymPptZYM48cIBKVu5z17W4M2jtrn1a3GJQ}{interactions extracted from WikiPathways version 20161210} & 2017-05-11 & & 10100 \\
48 & \href{http://purl.org/np/RAz6aGMobsbBYFWTp95n2RQOYZI8ZerInelEow0fsWGko}{interactions extracted from WikiPathways version 20170210} & 2017-05-11 & & 10375 \\
49 & \href{http://purl.org/np/RALEiGJ4h5ddSN92eL3AlewfhCxrdgeZ1YLhWjR_1wZFE}{interactions extracted from WikiPathways version 20170310} & 2017-05-11 & & 10371 \\
50 & \href{http://purl.org/np/RAd2XME41qX2G0_L8iyKZ5HQfzXc9iPoR4JKqlfxjr67I}{interactions extracted from WikiPathways version 20170410} & 2017-05-11 & & 10371 \\
51 & \href{http://purl.org/np/RApizOwYz0AfBo6y1TSmG_5hWCbXYCKVa2HspzrMEoTjk}{interactions extracted from WikiPathways version 20170510} & 2017-05-11 & & 10371 \\
52 & \href{http://purl.org/np/RAieZ49uxdKeXrqyDXvYC_fTnribtq9_fvsZdodCTJ-tk}{pathwayParticipation extracted from WikiPathways version 20161110} & 2017-05-11 & & 3830 \\
53 & \href{http://purl.org/np/RAlqhe3FmRcaPsQP9F6z6mlpc7QcS-87fbFYUmQm1zCVI}{pathwayParticipation extracted from WikiPathways version 20161210} & 2017-05-11 & & 3838 \\
54 & \href{http://purl.org/np/RAwx3ICvG1xAzLye1c2TCYKAf3EilnyBP1DCcIqdSlZLg}{pathwayParticipation extracted from WikiPathways version 20170210} & 2017-05-11 & & 3906 \\
55 & \href{http://purl.org/np/RAtKHwlPZQSYGG6pXxcvZ-lrRwRsiCxfpO48L5KEXn4jA}{pathwayParticipation extracted from WikiPathways version 20170310} & 2017-05-11 & & 3906 \\
56 & \href{http://purl.org/np/RA7H_R7RZRiJ14J3KTFD9IGzdnlWlp1kCJWlJWlTHseno}{pathwayParticipation extracted from WikiPathways version 20170410} & 2017-05-11 & & 3910 \\
57 & \href{http://purl.org/np/RAFyJrsxjjkkT2ALN2PkgQTI4YrHVfYgygpxqmxMhSMqA}{pathwayParticipation extracted from WikiPathways version 20170510} & 2017-05-11 & & 3910 \\
58 & \href{http://purl.org/np/RAkl7dPvlpTxXv3b5skIPdcseA0707M0Ab63FfBsFvEAc}{WikiPathways, incremental dataset, 20160610} & 2017-05-11 &2 & 9018 \\
59 & \href{http://purl.org/np/RAv1c3VBY0-0FSlVYsyXzqecgnd0V4zeeE83eDnRgwzKQ}{WikiPathways, incremental dataset, 20160710} & 2017-05-11 &2 & 10173 \\
60 & \href{http://purl.org/np/RAn72jhK8wPzZIbFLMNBEHk9s3JYRr2VEJkSji6vd151A}{WikiPathways, incremental dataset, 20160810} & 2017-05-11 &2 & 10123 \\
61 & \href{http://purl.org/np/RA2XpJikTwBSreEKut0jtQ52BAXV2Bulei-YS-9ya6CWI}{WikiPathways, incremental dataset, 20160910} & 2017-05-11 &2 & 10124 \\
62 & \href{http://purl.org/np/RAUv52zWreL97X4-s8ChdQ7yzma_TqXWiGL3oeISLqpQY}{WikiPathways, incremental dataset, 20161010} & 2017-05-11 &2 & 10127 \\
63 & \href{http://purl.org/np/RANmrwy03InvBPOTS31CDn-mMQBLS-CId_3ctqlgNs5b0}{WikiPathways, incremental dataset, 20161110} & 2017-05-11 &3 & 13958 \\
64 & \href{http://purl.org/np/RALSbi8hYBwO0LRScOREHn6wRecg0V87xdQ6fKlzZbh0s}{WikiPathways, incremental dataset, 20161210} & 2017-05-11 &3 & 13975 \\
65 & \href{http://purl.org/np/RAyZp9GEJJixTX91NqTHvJX4sxuXVqfIR3s453GxdTiUY}{WikiPathways, incremental dataset, 20170210} & 2017-05-11 &3 & 14323 \\
\end{tabular}
\end{table*}

\begin{table*}[tb]
\centering
\caption{Overview of nanopublication indexes (continued from Table \ref{tab:indexes1})}
\label{tab:indexes2}
\small
\begin{tabular}{rllrr}
\# & Index Title (some abbreviated) & Date & Sub & Size \\
\hline
66 & \href{http://purl.org/np/RAQoM3x3EezX6HL-FO48oLNv3xwHhSJjavnw2cDuMXDm8}{WikiPathways, incremental dataset, 20170310} & 2017-05-11 &3 & 14319 \\
67 & \href{http://purl.org/np/RAbZxrCNajdfj4BMq6WAixbB1hsshC8_hcBlmoaZ3clUU}{WikiPathways, incremental dataset, 20170410} & 2017-05-11 &3 & 14323 \\
68 & \href{http://purl.org/np/RAKz0OQ3Dq8dDWqF7SIY4TgYcZRX4d2TnmLUEbOwnaGmQ}{WikiPathways, incremental dataset, 20170510} & 2017-05-11 &3 & 14323 \\
69 & \href{http://purl.org/np/RAAGy9p2PABQk-MlGDF30myYrhJeDiqDpL9D5L82Knm3E}{\emph{(unnamed)}} & 2017-05-11 & & 83771 \\
70 & \href{http://purl.org/np/RAcf4tihZLL_aK81hwThIrNxjOhks4sEloBStEgzyR1tI}{\emph{(unnamed)}} & 2017-05-11 & & 4859 \\
71 & \href{http://purl.org/np/RAxMyDRaM8RmKGNiEe7dQPRUTuz616iI-N2T-H3MPYmXk}{\emph{(unnamed)}} & 2017-05-11 & & 18098 \\
72 & \href{http://purl.org/np/RAeyljAme-dwlpRSljxhq5Z2ga5SRRU3cs32Q1uBxtlXU}{Data for meta-analysis of polycystic kidney disease expression profiles} & 2017-05-18 & & 1657 \\
73 & \href{http://rdf.disgenet.org/nanopub-server/RAfh_E6xmcdxDifBCA5ojx061d4pfK6GHCdbiNqRxNou8}{DisGeNET v5.0.0.0} & 2017-10-17 & & 1469541 \\
74 & \href{http://purl.org/np/RAp9QBqvyPlvIoOpA_fCCVYHHlsInHIu2EmxOFQd7yRQg}{complexes extracted from WikiPathways version 20170610} & 2017-12-18 & & 42 \\
75 & \href{http://purl.org/np/RA-W56ZJP0U84frO_AnS_ZtyicjCf2IjNYK18bJ8FpZao}{interactions extracted from WikiPathways version 20170610} & 2017-12-18 & & 10371 \\
76 & \href{http://purl.org/np/RA9h-sX7g9x_R2sLq8gpyIwgyYpijvp4YIuACWdt3UXbQ}{pathwayParticipation extracted from WikiPathways version 20170610} & 2017-12-18 & & 3910 \\
77 & \href{http://purl.org/np/RA-OqlWv038E_6gEuB-7_Q0Z1vg1Riztes6wbO72K4T6Y}{WikiPathways, incremental dataset, 20170610} & 2017-12-18 &3 & 14323 \\
78 & \href{http://purl.org/np/RAWI1WaLMfpctXfhFSh0rvYrhm_EQ82a1YALToHk3cXqM}{complexes extracted from WikiPathways version 20170710} & 2017-12-18 & & 42 \\
79 & \href{http://purl.org/np/RAmEI2cMCUyH1WTNguP_M7NvsZoHhtosnoE45jj3AwprA}{interactions extracted from WikiPathways version 20170710} & 2017-12-18 & & 10371 \\
80 & \href{http://purl.org/np/RAdJFUnFeXykbkij-_0qLhdh7iVnM8_46JQihRmGTeR6E}{pathwayParticipation extracted from WikiPathways version 20170710} & 2017-12-18 & & 3910 \\
81 & \href{http://purl.org/np/RATjW4Rl_8gGY_SdlIYaaRuLiKMBYKtHZ78MoiuHDzwg8}{WikiPathways, incremental dataset, 20170710} & 2017-12-18 &3 & 14323 \\
82 & \href{http://purl.org/np/RArhG7E2zajatC9l5YfYJ97yrdvNn2gAJjyJV5G9X4xLA}{complexes extracted from WikiPathways version 20170810} & 2017-12-18 & & 43 \\
83 & \href{http://purl.org/np/RAGziCxT6VmT0P88oV1fojSV4uGTDN6GRMhWQtYivAm5Q}{interactions extracted from WikiPathways version 20170810} & 2017-12-18 & & 13208 \\
84 & \href{http://purl.org/np/RA6-Jweu-ZkTqtm1bSfncIJGZnxf_f3CNmeIE9aYa8CjE}{pathwayParticipation extracted from WikiPathways version 20170810} & 2017-12-18 & & 4362 \\
85 & \href{http://purl.org/np/RA0MYQBxAkmFTGzOUgmorKF-Z05E_n8snH1X3ZWNjmiEQ}{WikiPathways, incremental dataset, 20170810} & 2017-12-18 &3 & 17613 \\
86 & \href{http://purl.org/np/RATpzeUn6EUnsbbcjDjykIzMp12L46pryNmKb2fBIeI-A}{complexes extracted from WikiPathways version 20170910} & 2017-12-18 & & 43 \\
87 & \href{http://purl.org/np/RAKxH7Q3QEzfFNn0UTBZDXdS2zc4msoOHTKmHJ8uK0Es0}{interactions extracted from WikiPathways version 20170910} & 2017-12-18 & & 13080 \\
88 & \href{http://purl.org/np/RA0_EvKhRAp4c-nmRDAVnJFHjN66QpxzTEyuCsqN2mTic}{pathwayParticipation extracted from WikiPathways version 20170910} & 2017-12-18 & & 4364 \\
89 & \href{http://purl.org/np/RAl8UDvys8ZvCQZPiIKuH6vRws9iK79G-imHk2b4ks_BY}{WikiPathways, incremental dataset, 20170910} & 2017-12-18 &3 & 17487 \\
90 & \href{http://purl.org/np/RAnlfiUUQO_Maf9bB5dJaMDv59Mp8v2K3jFJNoJyuAP9k}{complexes extracted from WikiPathways version 20171010} & 2017-12-18 & & 43 \\
91 & \href{http://purl.org/np/RA8Z6vtMDkJwrfCwXTeCphiZcbtWcqTOVjOOqGEM5m8EY}{interactions extracted from WikiPathways version 20171010} & 2017-12-18 & & 12620 \\
92 & \href{http://purl.org/np/RAOPQkxpfN6Q0x1c9rPG4YE1qQpra823ANMdl9Cw3P4Cc}{pathwayParticipation extracted from WikiPathways version 20171010} & 2017-12-18 & & 4386 \\
93 & \href{http://purl.org/np/RATKvfPRl_biwB4p-35enMzZ-ssqEMWEJLIV2PA8ql-HY}{WikiPathways, incremental dataset, 20171010} & 2017-12-18 &3 & 17049 \\
94 & \href{http://purl.org/np/RAmV1dLULcQ1Y0XJ2a4XU8Um1seA_W6OLzM7F50qj5YQ4}{complexes extracted from WikiPathways version 20171116} & 2017-12-18 & & 44 \\
95 & \href{http://purl.org/np/RAEPIhZ2sqkfW2bksN7HGCnPsc-2c5Dl0i0hZijTHbO_I}{interactions extracted from WikiPathways version 20171116} & 2017-12-18 & & 12632 \\
96 & \href{http://purl.org/np/RA9PgElv4BBEaEFSBrrs_cNlpM6C00kanqjiZKYqHtdLg}{pathwayParticipation extracted from WikiPathways version 20171116} & 2017-12-18 & & 4403 \\
97 & \href{http://purl.org/np/RA02JkPYQWNKYOl1rFM-bAoW34J--0-WMabwzAiYUfs7E}{WikiPathways, incremental dataset, 20171116} & 2017-12-18 &3 & 17079 \\
98 & \href{http://purl.org/np/RAaeLg9HD-3KD_lG6iJwHXhH_nGQ-1o3DCZRRmzDY-at8}{complexes extracted from WikiPathways version 20171210} & 2017-12-18 & & 44 \\
99 & \href{http://purl.org/np/RAvqkqpEDxgu_CAx00IZSc8MYQqizkQaL4TsB72ThOCqQ}{interactions extracted from WikiPathways version 20171210} & 2017-12-18 & & 12638 \\
100 & \href{http://purl.org/np/RAjiRqNKwIGPRUinw1kgAr5_hLaTrnJqqWLMO1lMfpzYA}{pathwayParticipation extracted from WikiPathways version 20171210} & 2017-12-18 & & 4407 \\
101 & \href{http://purl.org/np/RA6QNVGZTuDN0g-zEKm02oLlqPO9JVz6cA4J760Bs0Zc4}{WikiPathways, incremental dataset, 20171210} & 2017-12-18 &3 & 17089 \\
102 & \href{http://purl.org/np/RANZaA3Io8uQjnyqmiIWuazXFdugH_hwbB2N5_LuF8bsY}{complexes extracted from WikiPathways version 20180110} & 2018-03-30 & & 44 \\
103 & \href{http://purl.org/np/RAviMsz7k0GorXIUUpz_70L1PjlXYYSugmQLOhL7I4_2w}{interactions extracted from WikiPathways version 20180110} & 2018-03-30 & & 12645 \\
104 & \href{http://purl.org/np/RAtx2TuK8OQBpdA0s4N0vbQtn7EnR5SJ9wHnqeCwiS66c}{pathwayParticipation extracted from WikiPathways version 20180110} & 2018-03-30 & & 4408 \\
105 & \href{http://purl.org/np/RA_AWcwm57iWtR5GbfhFAV2L2xsGMid8usbrtE-2h7Rzk}{WikiPathways, incremental dataset, 20180110} & 2018-03-30 &3 & 17097 \\
106 & \href{http://purl.org/np/RAMBAEXMXBFUJSO8jVsYdhqmys2VU-oqlMcxiJTzFrv8w}{complexes extracted from WikiPathways version 20180210} & 2018-03-30 & & 45 \\
107 & \href{http://purl.org/np/RAje8Xo6w9QRPOzaLLaOHvCPFT7frFf3WlJBZY9oui0_I}{interactions extracted from WikiPathways version 20180210} & 2018-03-30 & & 12683 \\
108 & \href{http://purl.org/np/RAt5TYmiahHGyJ3DOC1-Ys95tGb2Yn4EYwXDuNSpZ5Lqw}{pathwayParticipation extracted from WikiPathways version 20180210} & 2018-03-30 & & 4413 \\
109 & \href{http://purl.org/np/RAkIUxPxzVhUMNZ8up0ivmR3th6-qhFvf7LX6oJM78xws}{WikiPathways, incremental dataset, 20180210} & 2018-03-30 &3 & 17141 \\
110 & \href{http://purl.org/np/RA1AMnSjbML9T4WDOsTZu5UDhD7Mr8ExIvztGd4Y4uJfk}{complexes extracted from WikiPathways version 20180310} & 2018-03-30 & & 45 \\
111 & \href{http://purl.org/np/RA4pAKrybhUh4Y6qazLEitSZ7rXqcxUrGVezfmT4yhIdQ}{interactions extracted from WikiPathways version 20180310} & 2018-03-30 & & 12685 \\
112 & \href{http://purl.org/np/RAMjL9sevrzITxZdeB_tUqfRABp5PlTYhdiNcq_kKz440}{pathwayParticipation extracted from WikiPathways version 20180310} & 2018-03-30 & & 4417 \\
113 & \href{http://purl.org/np/RAxem_NvV-uMTtUXUePxcdib0xw90bT1TExLJQp-Z7cgA}{WikiPathways, incremental dataset, 20180310} & 2018-03-30 &3 & 17147 \\
114 & \href{http://purl.org/np/RAD3LTE5gKujkUmYPfY0yhxZHeYW5kwpF_f3YBxYoB7AA}{Core drug data extracted from Drugbank via Bio2RDF} & 2018-03-30 & & 7740 \\
115 & \href{http://purl.org/np/RAUgzHCC5Z0uH8v89Bddd9NMdTn7ctK7vZHpOve7YPaG4}{Drug-drug interaction data extracted from Drugbank via Bio2RDF} & 2018-03-30 & & 12104 \\
116 & \href{http://purl.org/np/RAMYp-s7FxlewSUzyh2LLSQbXnjngkuNXCjlIkDe5h0TY}{Food interaction data for drugs extracted from Drugbank via Bio2RDF} & 2018-03-30 & & 310 \\
117 & \href{http://purl.org/np/RAZUFR72KsiOtMWlxTSyGnZzQZylkBItGXDVe0ADmHUOs}{Drug target data extracted from Drugbank via Bio2RDF} & 2018-03-30 & & 4007 \\
118 & \href{http://purl.org/np/RAK_mkjiDrgvlttNWZHgBcRrpepiAiRgVBopYcKUZqUN4}{Drug target relations extracted from Drugbank via Bio2RDF} & 2018-03-30 & & 15107 \\
119 & \href{http://purl.org/np/RAbyBsPKDTg5OGzkHuk4GM47PQxb4uOtRluFpSaXZn-Rw}{Drug data extracted from Drugbank via Bio2RDF} & 2018-03-30 &5 & 39268 \\
120 & \href{http://purl.org/np/RAM-2ntLXzb2WY1VM_gTrVQDydc6cSVG0fMgB26HmIhnU}{Linked Drug-Drug Interactions (LIDDI) dataset v1.02} & 2018-03-30 & & 98085 \\
121 & \href{http://purl.org/np/RACoe1hCsz5r5mPs99Eggb_auVEeB-SoM6YTAfZrEDz-A}{Dataset of loose nanopublications} & 2018-04-04 & & 42 \\
122 & \href{http://purl.org/np/RAZqN5PLPHvTo0zvpdET8nLQ2FCCn2s3UGrcBun2ObSkQ}{Dataset of loose example nanopublications} & 2018-04-04 & & 183 \\
123 & \href{http://purl.org/np/RAf8SwAtE51Q8ZUn0f2VoinFPlxj5h9WIIxMSWjDCLHVw}{Nanopublications for the Avian Diet Database} & 2018-04-05 & & 25962 \\
124 & \href{http://purl.org/np/RA3evbIf5lh5ZM7rnX9CAvfmZtkVFw__KCXLByR5dQkbQ}{Nanopublications for Turfgrass Diseases} & 2018-04-05 & & 3430 \\
125 & \href{http://purl.org/np/RAO3tQt2Hg2rbbroMSZFyCKUi_uTxeeMjNijkC-4DKteo}{Nanopublications for iNaturalist.org Species Interaction Observations} & 2018-04-05 & & 29882 \\
126 & \href{http://purl.org/np/RAIW-kUqvdYwQgcj9leWpys3c758PR9JS4AxZbXWXCXKE}{Nanopublications for Catalogue of Afrotropical Bees} & 2018-04-05 & & 5267 \\
127 & \href{http://purl.org/np/RAdBmj3YeUUy6CwMTFZPPX2ezaQTOt5CqLMS331cJP2W0}{Nanopublications for Species Interaction Dataset by Raymond et al.} & 2018-04-05 & & 14961 \\
128 & \href{http://purl.org/np/RARgSSH931OncISJBf7NAzZAnCxg_416-pZZIBKhYz9Tg}{Nanopublication Collection of Global Biotic Interactions, Version 1} & 2018-04-05 &5 & 79502 \\
129 & \href{http://purl.org/np/RAWfIXHNHhEl9nlHbsuIsZuXFigxWpbw3o3m6YEecOERo}{Monogenic rare diseases: gene associations with specific literature references} & 2018-04-05 & & 4583 \\
\end{tabular}
\end{table*}

\bibliographystyle{splncs04}
\bibliography{references}

\end{document}